\documentclass[12pt]{article}
\usepackage{graphicx}
\usepackage[linktocpage=true,colorlinks=true,linkcolor=blue,citecolor=blue]{hyperref}
\usepackage{cite}
\usepackage[utf8]{inputenc}

\begin{document}
\def\beq{\begin{equation}}
\def\eeq{\end{equation}}
\newcommand{\comment}[1]{}

\begin{center}
{\Large \bf
Indirect measurement of triple-Higgs coupling \\ at an electron-positron 
collider \\{\vskip .2cm} with polarized beams
}

\bigskip
\bigskip

{\large
Saurabh D. Rindani$^{a,}$\footnote{saurabh@prl.res.in} and Balbeer
Singh$^{a,b,}$\footnote{balbeer@prl.res.in}}\\

\vskip .4cm
$^a${\it Theoretical Physics Division,
Physical Research Laboratory\\
Navrangpura, Ahmedabad 380 009, India}\\
\vskip .3cm
$^b${\it Indian Institute of Technology Gandhinagar\\
 Gandhinagar 382 355,
Gujarat, India}

\bigskip
\bigskip

{\bf Abstract}

\end{center}

{\footnotesize
We examine the possibility of using single-Higgs production at an 
$e^+e^-$ collider with polarized beams to measure, or constrain,
indirectly a possible anomalous triple-Higgs coupling, which can
contribute to the process via one-loop diagrams. In the
dominant process $e^+e^- \to ZH$, longitudinally polarized beams can
lead to an improvement in the cross section by 50\% for $e^-$ and $e^+$ 
polarizations of $-0.8$ and $+0.3$, respectively. This corresponds to an
improvement in the sensitivity to the triple-Higgs coupling of about
18\% for a centre-of-mass energy of 250 GeV and an integrated luminosity
of 2 ab$^{-1}$, making a strong case of beam polarization. This also
implies that with polarized beams, the luminosity needed to get a
particular sensitivity is less by about 33\% as compared to that
needed with unpolarized beams. Even when
only the $e^-$ beam is polarized $-0.8$, the improvement in the sensitivity 
is about 8\%.  We also study the effect of longitudinal 
beam polarization on the sensitivity to the triple-Higgs coupling of
Higgs production 
through the subdominant process  $e^+e^- \to H \nu \bar
\nu$ occurring through $WW$ fusion.  

%Keywords:Higgs boson, triple-Higgs coupling, electron-positron
%colliders, polarized beams
}

\bigskip

\section {  Introduction} 
Experiments at the Large Hadron Collider (LHC) since the discovery of the
Higgs boson have been steadily enhancing the accuracy in the measurement
of its couplings to various standard model (SM) particles and the
results show good conformity with theoretical predictions within the SM.
An exception to this trend of high accuracy measurements is the scalar
self-couplings, which as yet have not been determined with very good
precision. The scalar self-couplings correspond to $\lambda_3$ and
$\lambda_4$ in the following terms in the scalar Lagrangian
\beq\label{3H}
{\cal L}_{3H} = - \lambda_3 H^3,
\eeq
\beq\label{4H}
{\cal L}_{4H} = - \lambda_4 H^4.
\eeq
In the SM, these couplings are related to the physical Higgs mass $m_H$
and the scalar vacuum expectation value $v$ by
\beq\label{SMlambdas}
\lambda_3 = \lambda_4 v ; \;\; \lambda_4 = m_H^2/(2 v^2).
\eeq
It will be the task of future experiments at the LHC as well as at the
proposed lepton colliders to determine $\lambda_3$ and $\lambda_4$ with
greater precision and check the SM relations of eq. (\ref{SMlambdas}).
It is of course possible that the underlying theory is not the SM, but an
extension of the SM. In that case, the above couplings would be
the couplings in an effective theory, and they may not obey the
relations (\ref{SMlambdas}). Deviations of these couplings from
their SM values have been discussed in the context of the standard model
effective field theory (EFT), 
where effective interactions induced by new physics
are written in terms of higher-dimensional operators, suppressed by a
high-energy scale, the effective theory presumed to be valid at energies
much lower than this scale. Thus, for example, ${\cal L}_{3H}$ would get
a contribution from a dimension-six operator, $-\lambda (H^\dag H)^3$,
see for example, \cite{Degrassi:2016wml}.

A determination of the triple-Higgs coupling $\lambda_3$ can be
carried out through a process where two or more Higgs bosons are
produced. First of all, such a process needs a high centre-of-mass
(c.m.) energy for the interacting particles, partons in the case
of hadron colliders, or leptons in the case of lepton colliders.
Moreover, it has been found that in the SM \cite{glover}, there is
destructive interference between the one-loop diagrams contributing to the 
process $ gg \to HH$,
making the total cross section extremely small. Thus, the accuracy of
the determination of $\lambda_3$ is low.

A suggestion was made by McCullough \cite{McCullough:2013rea} 
that the triple-Higgs
coupling could be measured through its contribution in one-loop diagrams
in single-Higgs production. The process considered in \cite{McCullough:2013rea}
was $e^+e^- \to ZH$, in which, making an assumption that only
$\lambda_3$ deviates from its SM value,
\beq\label{kappa}
\lambda_3 = \lambda^{\mathrm SM}v (1 + \kappa),
\eeq
it would be possible to put a limit on the fractional deviation $\kappa$
which is of the order of 28\% at $e^+e^-$ c.m. energy of 240 GeV, with
an integrated luminosity of 10 ab$^{-1}$, expected to be available at
TLEP (currently known as FCC-ee) \cite{TLEP}.
This estimate is based on the assumption, made
for the sake of concreteness, that
there are no other contributions to an effective $ZZH$
vertex.

The proposal to construct a linear $e^+e^-$ collider, the so-called
International Linear Collider (ILC), in Japan may reach a decision in the
near future. We concentrate mainly on the ILC in this work, though our
discussion is also important in the context of other $e^+e^-$ colliders.
Since in the initial run the ILC would collect 2 ab$^{-1}$,
rather than 10 ab$^{-1}$ being anticipated for the
FCC-ee, it seems advisable to look for some means of enhancing the 
sensitivity of the ILC.
We therefore turn to beam polarization for the purpose.
We have examined to what extent the sensitivity of the
single-Higgs production processes to the triple-Higgs coupling can be
improved by the use of longitudinal beam polarization. 
It is well-known that suitable longitudinal 
polarization could lead to a larger cross section for $ZH$ production, and
may also help in suppressing background.
We find that beam polarization unfortunately does not improve the
relative contribution of the the triple-Higgs coupling. 
However, it does improve
the sensitivity because of the increase in statistics. A similar
improvement is also found for single Higgs production via $WW$ fusion,
though the dependence on beam polarization is somewhat different.
Thus, to achieve the same sensitivity for $\kappa$, a collider with
polarized beams would need much less luminosity as compared to one
without beam polarization. This makes a good case for beam
polarization at future $e^+e^-$ colliders. We have systematically
investigated this issue quantitatively, making comparisons of the results
that can be achieved without polarization and 
using electron and/or positron beam polarization.

Here we give a quantitative summary of our results. We assume that for a 
c.m. energy of 250 GeV, an
integrated luminosity of 2 ab$^{-1}$ would be available with polarized
beams. The degree of polarization assumed is $-0.8$ for electrons and
$+0.3$ for positrons. We have assumed that, as estimated in 
\cite{Fujii:2018mli, Fujii:2017vwa}, a precision of about 0.9\%
could be possible in the measurement of the cross section for $e^+e^-
\to ZH$. We then find that the accuracy of determination of the parameter 
$\kappa$ that could
be obtained from the measurement of the cross section with polarized beams 
is about 57\% as compared to 70\% in the absence of polarization,
an enhancement in sensitivity of about 18\%. To put it differently, at a
collider like the FCC-ee, if operating at 240 GeV with polarized beams, 
the same sensitivity
as anticipated in ref. \cite{McCullough:2013rea} could be achieved with
an integrated luminosity of about 6.66 ab$^{-1}$. We also found that the 
sign-inverted combination of polarizations, {\it viz.}, $+0.8$ and $-0.3$
respectively for electrons and positrons, gives worse sensitivity
as compared to the unpolarized case.

The plan of the remaining paper is as follows. In the next section, 
we summarize the different approaches
to the measurement of the Higgs self-coupling at different colliders.
In section 3, we present the formalism, including analytical expressions, for
calculating the effect of the tree-level and the loop contribution to the
single-Higgs production processes. The effect of polarization is discussed in
Section 4. After presenting the numerical results in Section 5, the last
section (6) contains our conclusions and a discussion of the results.

\section{Various methods for measurement of Higgs self-coupling}

We discuss here in brief various methods suggested for the  measurement 
of the Higgs self-coupling.

Indirect constraints can be obtained from electroweak precision
measurements. A bound of $-15.0<\kappa < 16.4$ was obtained
\cite{Kribs:2017znd, Degrassi:2017ucl} by this method.

As mentioned earlier, the dominant approach to the measurement of the Higgs
self-coupling is through Higgs pair production.
The first possibility is Higgs pair production at the LHC, and then at the 
future high luminosity run of the LHC (HL-LHC).
It is expected that at
HL-LHC, with a c.m. energy of 14
TeV and an integrated luminosity of 3 ab$^{-1}$, with sophisticated jet
substructure techniques, a limit of $\lambda_3 < 1.2 \lambda_3^{\rm SM}$
may be set \cite{deLima:2014dta}. 
Adding more Higgs decay channels, this accuracy might be
improved, see, for example, \cite{Kim:2018uty}.

A future hadron collider at 100 TeV may probe the coupling, 
using the $b\bar
b\gamma\gamma$ channel for HH decay, to an accuracy of 30\%
\cite{Azatov:2015oxa}. {For a discussion at a 100 TeV FCC-hh collider in the $t\bar{t}hh$ channel see Ref.\cite{Banerjee:2019jys}.}

A global fit which includes several inputs including that from
double-Higgs production 
was shown to enable limits on $\lambda_3/\lambda_3^{\rm
SM}$ in the interval [0.1,2.3] 
\cite{DiVita:2017eyz}.
A recent  work \cite{Goncalves:2018qas} proposes restricting to certain
kinematic regions to improve the accuracy of measurement at the LHC. 

There have been various suggestions for the
construction of $e^+e^-$ colliders with c.m. energy ranging from a few
hundred GeV to a few TeV. After the discovery of the Higgs boson with
mass of about 125 GeV, the dominant suggestion is to construct a linear
collider, the ILC, which would
first operate at a c.m. energy of 250 GeV, enabling precise measurement
of Higgs properties, through an abundant production of a $ZH$ final state. 
There have also been other proposals, as for example, the Compact Linear
Collider (CLIC), the Future Circular Collider (FCC-ee) and Circular
Electron Positron Collider (CEPC) where electron and positron beams
would be collided, providing a clean environment to study couplings of
SM particles, and possibly look for new physics, if any. 

Double Higgs production can also be used to measure $\lambda_3$
at an $e^+e^-$ collider.
For example, $\lambda_3$ would be determined with an accuracy
of 83\% at the proposed International Linear Collider (ILC) at c.m.
energy of 500 GeV, improving to 21\% and 13\% with luminosity
and energy upgrades
\cite{Baer, Asner, Fujii:2017vwa}.
%{\color{red} In a recent study in Ref.\cite{{Roloff:2019crr}}  
%it is shown that with an integrated luminocity of 5 ab$^{-1}$ and 
%$\sqrt{s}=3$TeV, CLIC will be able to measure trilinear higgs coupling 
%with an uncertainty of $-7$\% 
%and 11\% with C.L. 68\%.}

Some other recent studies on Higgs coupling measurements at electron-positron 
colliders, though not necessarily limited to triple-Higgs couplings,
are 
\cite{Maltoni:2018ttu}-
%Barklow:2017awn, DiVita:2017vrr,
%epluseminus, deBlas:2019rxi,deBlas:2018mhx,Roloff:2018dqu,Chiu:2017yrx,
%Baglio:2019gmz,Roloff:2019crr,
%An:2018dwb,Bambade:2019fyw, Nakamura:2018bli, 
\cite{Kumar:2019bmk}.
We will discuss the most relevant ones towards the end.

As mentioned earlier, single Higgs production in association with $Z$
at an $e^+e^-$ collider was suggested by McCullough \cite{McCullough:2013rea}
as an indirect measurement of $\lambda_3$ through loop contributions.
Following McCullough's suggestion, there have been proposals to measure
$\kappa$ in various single-Higgs production processes at the LHC
~\cite{bizon, Degrassi:2016wml, Gorbahn:2016uoy, Maltoni:2017ims} and
at $e^+e^-$ colliders 
\cite{Barklow:2017awn, DiVita:2017vrr, Maltoni:2018ttu}. For example,
the Run I single-Higgs data at the LHC
leads to $-13.6 \leq \lambda_3/\lambda_3^{\rm SM} \leq 16.9$
\cite{bizon}.

Single Higgs production has also been proposed at the future LHeC collider
as a possibility for the measurement of $\lambda_3$ 
\cite{Li:2019jba}. The best limits anticipated in this paper on $\kappa$
are [-0.10, 4.07] for an integrated luminosity of 3 ab$^{-1}$ 
proton and electron  beam energies of 7 TeV and 60 GeV, respectively.

%{\color{red}
HL-LHC would eventually be able to determine $\kappa$
to a good degree of precision, of the order of 50\% at 68\% confidence level
\cite{Cepeda:2019klc, deBlas:2019rxi}.
 However, it is expected that results from an $e^+e^-$
collider would be able to improve on this accuracy, especially when combined
with the HL-LHC results.
%}.

We now review the current limits on the Higgs self-coupling.
Current constraints from the analysis of the $\sqrt{s}=8$ TeV (Run I) 
and the $\sqrt{s}=13$ TeV (Run II) data from the LHC 
 are weak.
Direct searches constrain $\lambda_3$ to $-14.5 \leq
\lambda_3/\lambda_3^{\rm SM} \leq 19.1$ (Refs. \cite{Gorbahn:2016uoy} and
\cite{Aad:2015xja}) from Run I data.
%and 
%$-8.4 \leq
%\lambda_3/\lambda_3^{\rm SM} \leq 13.4$ \cite{bizon, ATLAS:2016ixk} from
%Run II data.
From the Run II data, using $\gamma\gamma b \bar b$ final state, ATLAS has
obtained a limit of $-8.2 < \lambda_3/\lambda_3^{\rm SM} < 13.2$ 
\cite{Aaboud:2018ftw}, whereas the corresponding
CMS limit is $-11 < \lambda_3/\lambda_3^{\rm SM} < 17 $ 
\cite{Sirunyan:2018iwt}. ATLAS has also studied $b\bar bb\bar b$ final
state, and reports a limit of $\lambda_3/\lambda_3^{\rm SM} < 17 $ 
\cite{Aaboud:2018knk}. {In Ref.{\cite{Aad:2019uzh}},  a recent study of Higgs boson pair decaying into various channels has constrained the ratio $-5 < \lambda_3/\lambda_3^{\rm SM} < 12 $ with 95\% C.L.}

In this work we pursue further the possibility of measuring 
$\kappa$ at
future $e^+e^-$ colliders. 
In the context
of $e^+e^-$ colliders, particularly the ILC, the possibility of
utilizing beam polarization and its advantages have received much
attention. Suitable longitudinal beam polarization could help 
in improving the sensitivity for
many different processes and suppressing unwanted background 
\cite{MoortgatPick:2005cw,Fujii:2018mli}.
The use of polarization in the context of Higgs properties is also
discussed in \cite{Durieux:2017rsg,Barklow:2017suo}. 
It is expected that at the ILC, polarizations of 80\% and 30\% would be
possible respectively for electron and positron beams for c.m. energy of
250 GeV \cite{Fujii:2018mli}. 
We therefore proceed in the next section 
with a calculation of the loop contribution to 
the cross section for $e^+e^- \to ZH$, and assume this combination of
polarizations to estimate the sensitivity of determination of $\lambda_3$
in the process.

\begin{figure}[htb]
\centering
\includegraphics[width=6cm]{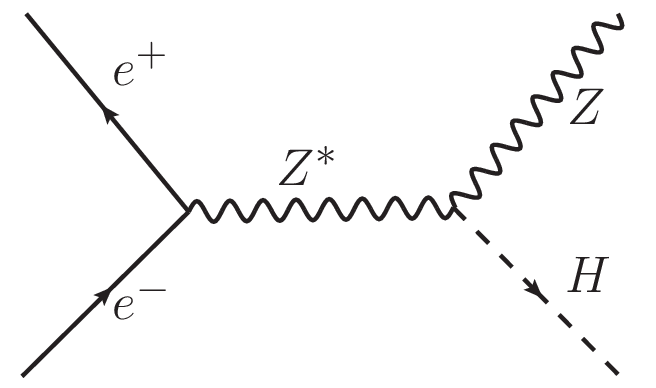}
\caption{Tree diagram for the process $e^+e^- \to ZH$.}\label{treezh}
\end{figure}
\begin{figure}[htb]
\centering
\includegraphics[width=6cm]{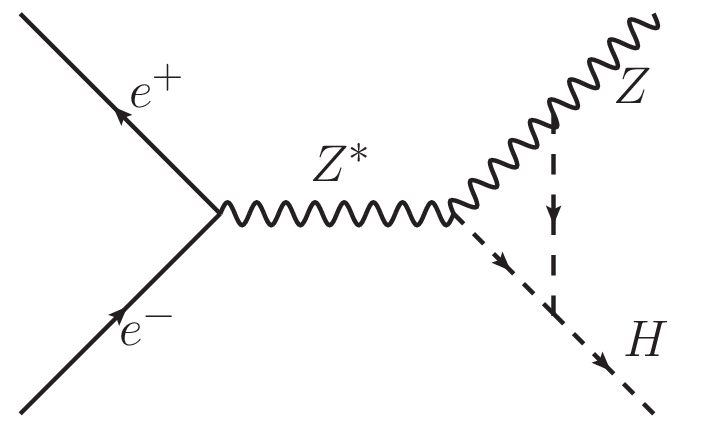}
\includegraphics[width=6cm]{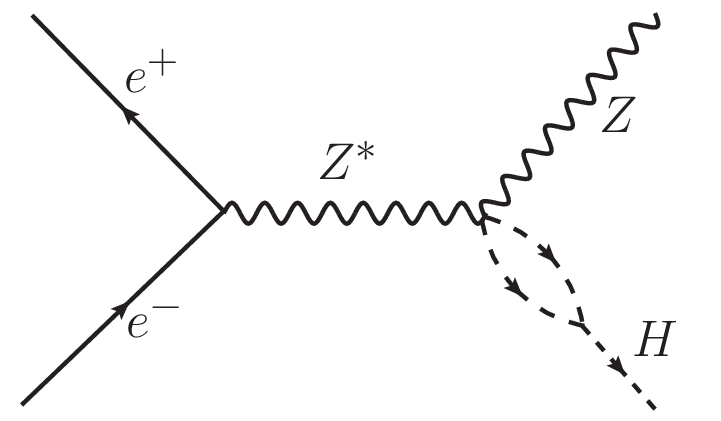}
\caption{One-loop diagrams for the process $e^+e^- \to ZH$ which include
the triple-Higgs coupling.}\label{loopzh}
\end{figure}
\section{Formalism} Here we briefly describe the formalism for
the
determination of $\lambda_3$ from single-Higgs production at non-leading
order. The calculation
of the effect of the triple-Higgs coupling at one-loop level follows
the work in refs. \cite{McCullough:2013rea,bizon}. 

The Feynman diagrams for the process $e^+e^- \to ZH$ are shown in Figs.
\ref{treezh} and \ref{loopzh}. The former shows the tree-level diagram,
whereas the latter shows the one-loop
diagrams which have contribution from the triple-Higgs coupling.

\begin{figure}[ht]
\centering
\includegraphics[width=5cm]{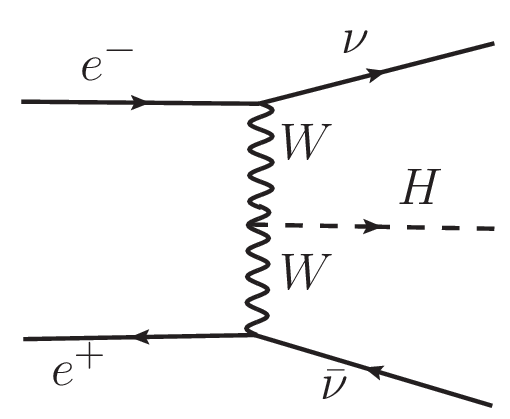} 
\caption{Tree diagram for the process $e^+e^- \to H\nu\bar\nu$ occurring
through $WW$ fusion.}\label{treeww}
\end{figure}

\begin{figure}[ht]
\centering
\includegraphics[width=5.8cm]{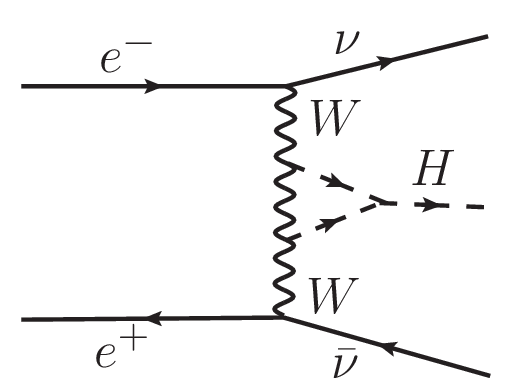}
\includegraphics[width=6cm]{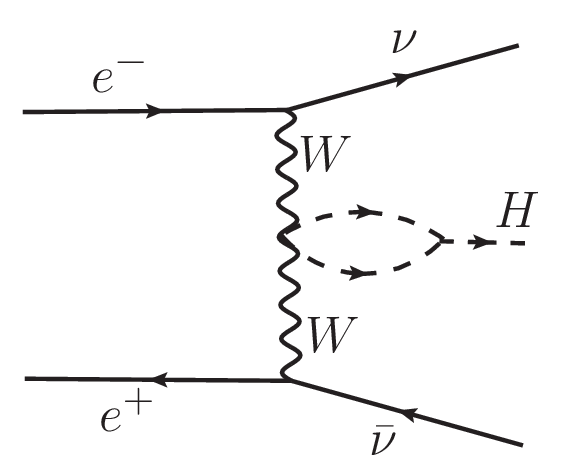}
\caption{One-loop diagrams for the process $e^+e^- \to H\nu\bar \nu$
occurring through $WW$ fusion which include the triple-Higgs
coupling.}\label{loopww}
\end{figure}

The Feynman diagrams for the $WW$-fusion process $e^+e^- \to H\nu\bar\nu$ 
are shown in Figs.
\ref{treeww} and \ref{loopww}. The former shows the tree-level diagram,
whereas the latter shows the one-loop
diagrams which have contribution from the triple-Higgs coupling.

The process $e^+e^- \to ZH$ as well as the $WW$ fusion contribution to 
$e^+e^- \to \nu \bar \nu H$ involve a $VV^*H$ vertex, where $V = Z$ in
the former case, and $V=W$ in the latter. The $VV^*H$ vertex can be
written as
\beq\label{VVH}
\Gamma^{\mathrm VVH}_{\mu\nu} = g_Vm_V[(1+{\cal F}_1)g_{\mu\nu}
        + {\cal F}_2k_{1\nu}k_{2\mu}],
\eeq
where $k_1$, $k_2$ are the momenta (assumed directed inwards) 
of the gauge bosons carrying the respective
polarization indices $\mu$, $\nu$. This form assumes that the gauge
bosons
are either on-shell, satisfying $k_{i\mu}\epsilon^{\mu}(k_i) = 0$ (i=1,2), or
couple to a conserved current, so that the terms with $k_{1\mu}$ or
$k_{2\nu}$ can be dropped. Here $m_V$ is the mass of the gauge boson,
$g_W$ is the weak coupling and $g_Z = g_W/\cos\theta_W$, $\theta_W$ being
the weak mixing angle. The quantities ${\cal F}_{1,2}$ are functions of
bilinear invariants constructed from the momenta.

Isolating the
contribution of the triple-Higgs coupling $\lambda_3$, the form factors
${\cal F}_{1,2}$ for the two processes of $Z^* \to ZH$ and $W^{+*}W^{-*}
\to H$ can be written at one-loop order in terms of the
Passarino-Veltman (PV) functions \cite{PV} as follows.
\beq\label{F1}
{\cal F}_1 ( k_1^2, k_2^2 ) = \frac{\lambda^{\rm SM}(1+ \kappa ) }{(4 \pi)^2}
        \left( - 3B_0 -12 (m_V^2 C_0 - C_{00} - \frac{9 m_H^2}{2} (\kappa
        +1))B'_0 \right),
\eeq
\beq\label{F2}
{\cal F}_2 (k_1^2, k_2^2 ) = \frac{\lambda^{\rm SM}(1 + \kappa ) }{(4 \pi)^2}
        12(C_1 + C_{11} + C_{12}).
\eeq
For the process $W^{+*}W^{-*} \to H$, the arguments of the PV functions
are 
\beq\label{PVWW}
B_0 \equiv B_0 (m_H^2, m_H^2, m_H^2), \;\; C_0 \equiv
C_0(m_H^2,k_1^2,k_2^2,m_H^2,m_H^2,m_W^2),
\eeq 
and analogously for the functions $B'_0$ and the tensor coefficients
$C_1, C_{11}$ and $C_{12}$.
For the process $Z^* \to ZH$, the arguments of the PV functions are
\beq\label{PVZH}
B_0 \equiv B_0 (m_H^2, m_H^2, m_H^2), \;\; C_0 \equiv
C_0(m_H^2,s,m_Z^2,m_H^2,m_H^2,m_Z^2),
\eeq
and analogously for the functions $B'_0$ and the tensor coefficients
$C_1, C_{11}$ and $C_{12}$.

The above expressions are evaluated to first order in the parameter
$\kappa$ for consistency, as there would be higher-loop contributions at
order $\kappa^2$ which are not being included. 
We use the package LoopTools \cite{looptools} to evaluate the PV
integrals.

In a realistic situation, there would be contribution at
tree level from an anomalous $ZZH$ vertex. This beyond standard model 
(BSM) contribution
would be model dependent, depending on the anomalous couplings. 
In a specific model, as for example, a two Higgs doublet model,
the tree-level $ZZH$ vertex of the SM would be modified by a 
known constant, determined by certain mixing angles in the theory.
In an effective field theory approach, the anomalous contribution
would be unknown, though the higher-dimensional operator would be
suppressed by powers of the cut-off scale.
Thus, a way 
should be found of extracting the loop-level contribution, 
separating the tree level BSM contribution. The extraction of $\kappa$
is possible either from the cross section using the
different energy dependences of the two contributions
\cite{McCullough:2013rea}, or from the differential cross section, 
using details of $Z$ decay distributions.
{It has also been pointed out that anomalous coupling of the
Higgs to a top-quark pair can affect the determination of the triple-Higgs
coupling, though this pollution is small at low energies
\cite{Shen:2015pha}.
}

In the next section we evaluate the cross section in the presence of beam
polarization, including the contribution of $\kappa$ at the one-loop level,
which was just discussed.

\section {Polarized cross section} 
The differential cross section for the process $e^+e^- \to ZH$ with
longitudinally polarized beams may be written as \cite{Rindani:2009pb}
\beq\label{diffcrossL}
\frac{d\sigma_L}{d\Omega} = ( 1 - P_L \bar P_L)[A_L + B_L \sin^2\theta
],
\eeq
where $\theta$ is the angle between $Z$ and $e^-$ directions, $P_L$, $\bar P_L$ are the degrees of longitudinal polarizations of
$e^-$, $e^+$ beams, and $A_L$ and $B_L$ can be split into their
respective tree-level SM
contributions $A_L^{\rm SM}$ and $B_L^{\rm SM}$ and the extra
contributions $\Delta A_L$ and $\Delta B_L$  
coming from the one-loop trilinear Higgs couplings.
The expressions for these are as follows. 
\beq\label{AL}
A_L = A_L^{\rm SM} + \Delta A_L,
\eeq
\beq\label{BL}
B_L = B_L^{\rm SM} + \Delta B_L,
\eeq
where
\beq
A_L^{\rm SM} = B_L^{\rm SM} \frac{2 m_Z^2}{\vert \vec q \vert^2}
	= (g_V^2 + g_A^2 - 2 g_V g_A P_L^{\rm eff})K^{\rm SM},
\eeq
\beq
K_{\rm SM} = \frac{\alpha^2\vert \vec q \vert }{2\sqrt{s}\sin^4
2\theta_W } \,\frac{m_Z^2}{(s-m_Z^2)^2},
\eeq
$q$ being the momentum of the $Z$, $g_{V,A}$ the SM couplings
of the $Z$ to $e^+e^-$,
\beq
g_V = -1 + 4 \sin^2\theta_W, \; g_A = -1,
\eeq
and
\beq
P_L^{\rm eff} = \frac {P_L - \bar P_L }{ 1 - P_L \bar P_L}.
\eeq
The expressions for the contributions from anomalous triple-Higgs couplings are 
\beq\label{DeltaAL}
\Delta A_L = 2 {\cal F}_1  (g_V^2 + g_A^2 - 2 g_V g_A P_L^{\rm
eff})K^{\rm SM},
\eeq
\beq\label{DeltaBL}
\Delta B_L = 2 \left( {\cal F}_1  + {\cal F}_2 \sqrt{s} q^0
\right)\frac{\vert q \vert^2}{2m_Z^2} (g_V^2 + g_A^2 - 2 g_V g_A P_L^{\rm
eff})K^{\rm SM}.
\eeq
The partial cross section, with a cut-off $\theta_0$ on forward and backward
directions is obtained by integrating $\theta$ between limits $\theta_0$
and $\pi - \theta_0$. The corresponding expression is
\beq\label{partialCSL}
\sigma_L(\theta_0) = ( 1 - P_L \bar P_L) 4\pi \cos\theta_0 [ A_L + 
\left( 1- \frac{1}{3} \cos^2\theta_0\right) B_L ].
\eeq
We see from the above equations that the beam polarization dependence 
of the SM contribution as well of the contribution of the triple-Higgs coupling 
to the cross section is through the same factor of $( 1 - P_L \bar P_L)(g_V^2 + g_A^2 - 2 g_V
g_A P_L^{\rm eff})K^{\rm SM}$. This is also true for the differential
cross section.

The differential cross section for the process $e^+ e^- \to H \nu \bar
\nu$ has two contributions, one from the process $e^+e^- \to ZH$,
with $Z$ decaying into $\nu \bar \nu$, and the other from the $WW$
fusion process $e^+ e^- \to W^+ \bar \nu W^- \nu \to H \nu \bar \nu$.
There is also a small interference term between the amplitudes for
these two mechanisms, which we neglect. The contribution from $ZH$
production followed by $Z \to \nu \bar \nu$ is the same as that obtained
for $ZH$ production earlier here, multiplied by the branching ratio for
$Z \nu \bar \nu$, which is about 20\%. The matrix element squared for
the $WW$ fusion process summed over final polarizations may be written as
\beq\label{wwfusion}
\begin{array}{rcl}
\Sigma \vert M \vert ^2& =& \displaystyle \frac{g_W^6 m_W^2 (1 - P_L) (1 + \bar
P_L)} { (q_1^2-m_W^2)^2 (q_2^2 - m_W^2)^2 } \left[ (1 + {\cal F}_1)^2
p_1\cdot p_4 p_2\cdot p_3 \right. \\ 
 && \left. +  (1 + {\cal F}_1){\cal F}_2 \left\{ s (
p_1\cdot p_4 p_2\cdot p_3 - \frac{1}{2} p_1\cdot p_4 p_3\cdot p_4 - \frac{1}{2}
         p_2\cdot p_3p_3\cdot p_4)\right. \right.\\ && \left. \left. + 
p_1\cdot p_4 p_2\cdot p_3 (2 p_3\cdot p_4
-p_2\cdot p_3 - p_1\cdot p_4 ) \right. \right. \\&&\left. \left. + p_1\cdot p_3 p_2\cdot p_4 (p_1\cdot p_4
+ p_2 \cdot p_3) \right\} \right].
\end{array}
\eeq

\section{Numerical Results} For our numerical calculations, we use 
$m_W = 80.385$ GeV, $m_Z = 91.1876$ GeV, $g_W =
\sqrt{8m_W^2(G_F/\sqrt{2})}$, $\sin^2\theta_W = 0.22$ and $m_H = 125$ GeV. 
We assume electron longitudinal polarization of $P_L = -0.8$, and
positron longitudinal polarization $\bar P_L = + 0.3$. 
For our analysis of the sensitivity, we assume a modest
integrated luminosity of 2 ab$^{-1}$, which for simplicity is taken to
be the same for unpolarized beams as well as all polarization
combinations. The efficiency of measurement of the final state is taken
from earlier works \cite{Asner,Fujii:2017vwa,Fujii:2018mli} to be 0.9\% at 
$\sqrt{s}=250$ GeV, and appropriately scaled for other energies. 
It may be mentioned that the efficiencies expected at other colliders
are 0.4\% at FCC-ee for $\sqrt{s}=240$ GeV and luminosity 10 ab$^{-1}$
\cite{TLEP}, 
3.8\% at CLIC for $\sqrt{s}=350$~GeV and luminosity 500 fb$^{-1}$
\cite{Abramowicz:2016zbo},
and 0.5\% at CEPC for $\sqrt{s}=240$ or 250 GeV and luminosity 5.6 ab$^{-1}$
\cite{An:2018dwb}. 

Taking up the case of longitudinal polarization for the process $e^+e^-
\to ZH$ first, we present in
Table 1 our
results for the polarization dependence of the cross section $\sigma_L$, 
as well as the 1 $\sigma$ limit that can be obtained on $\kappa$ 
from the polarized cross section for various values of c.m. energy.
We also present the fractional change $\delta\sigma_L$ in the cross section 
to the parameter $\kappa$ in
the form of $\delta\sigma_L /\sigma_L$ for unit change  in
$\kappa$, following ref. \cite{McCullough:2013rea}.
As mentioned earlier, this quantity does not change with polarization
for a given $\sqrt{s}$. This observation has also been made in
\cite{DiVita:2017vrr}, and follows from the fact mentioned earlier that
the polarization dependence of the cross section at tree level and at
one-loop in the triple-Higgs couplings is identical, both arising from
a single $Z$ at the $e^+e^-$ vertex \cite{Rindani:2009pb} .
In this case, we assume the cut-off on the production angle of the $Z$
to be zero, since it was found that the sensitivity worsens with increasing
cut-off angle. A negative sign in the table signifies that the cross
section decreases with increasing $\kappa$.

\begin{table}[htb]
\centering
\begin{tabular}{ccccrr}
\hline
$\sqrt{s}$ & $P_L$ & $\bar P_L$ &$\sigma_L$ (fb)&
$\frac{\delta\sigma}{\sigma}/\kappa$ & $\kappa_\mathrm{lim}$ (\%) \\ 
\hline
%250 & 0 & 0 & 248.9 & 1.264 &  70.8   \\
%    & $-0.8$ & 0 & 280.5 & 1.264 & 66.7   \\
%    & $-0.8$ & 0.3 & 352.1 & 1.196 & 59.5   \\
%350 & 0 & 0 & 131.6 & 0.278 &  322   \\
%    & $-0.8$ & 0 & 148.4 & 0.278 & 303   \\
%    & $-0.8$ & 0.3 & 186.2 & 0.243 & 271   \\
%500 & 0 & 0 &  58.0 &$-0.205$ & $- 437$   \\
%    & $-0.8$ & 0 &  65.5 &$-0.205$&$-412$ \\
%    & $-0.8$ & 0.3 &  82.0 &$-0.145$&$-368$\\
%1000 & 0 & 0 &  13.0 &$-0.431$ & $- 208$   \\
%    & $-0.8$ & 0 &  14.6 &$-0.431$&$-196$ \\
%    & $-0.8$ & 0.3 &  18.3 &$-0.288$&$-175 $\\
 250 & 0 & 0 & 242 & 1.278 &  70.0   \\
     & $-0.8$ & 0 & 288 & 1.278 & 64.2   \\
     & $-0.8$ & $+0.3$ & 364 & 1.278 & 57.2   \\
 350 & 0 & 0 & 129 & 0.284 &  315   \\
     & $-0.8$ & 0 & 153 & 0.284 & 289   \\
     & $-0.8$ & $+0.3$ & 193 & 0.284 & 257   \\
 500 & 0 & 0 &  56.9 &$-0.203$ & $- 440$   \\
     & $-0.8$ & 0 &  67.6 &$-0.203$&$-403$ \\
     & $-0.8$ & $+0.3$ &  85.3 &$-0.203$&$-359 $\\
 1000 & 0 & 0 &  12.7 &$-0.433$ & $- 206$   \\
     & $-0.8$ & 0 &  15.1 &$-0.433$&$-189$ \\
     & $-0.8$ & $+0.3$ &  19.1 &$-0.433$&$-169 $\\
\hline
\end{tabular}
\caption{The cross section $\sigma_L$ for $ZH$ production, 
the fractional change in the cross section $\sigma_L$ for unit
value of $\kappa$ and the sensitivity to $\kappa$ for various c.m.
energies $\sqrt{s}$ and combinations of $e^-$ and $e^+$ longitudinal
polarizations $P_L$ and $\bar P_L$, respectively. An integrated
luminosity of 2 ab$^{-1}$ is assumed.}
\end{table}
The points worth noting are that there is an improvement of sensitivity
in the measurement of $\kappa$ 
when both electron and positron beams are polarized.
The sensitivity improves even when only the electron beam is
polarized.
The increase in sensitivity is related to increase in the cross section by 
the factor $(1-P_L\bar P_L)(1-2g_Vg_A P_L^{\rm eff}/(g_V^2+g_V^2))$, 
which for the best case is approximately 0.5.
It can be checked that for the case of electron and positron polarizations
being respectively $+0.8$ and $-0.3$, this factor is less than 1, and
corresponds to degradation of the sensitivity. 

For the process $e^+e^- \to H\nu\bar \nu$ which includes the
Higgsstrahlung process followed by $H$ decay into 3 species of neutrino
pairs, as well as the $WW$ fusion process, the two contributions have to
be added. There is also a small interference term, which is neglected
for present purposes. Moreover, we make a simplifying assumption that
the accuracy of measurement is the same as in the case of $e^+e^- \to
ZH$ process. Our results are presented in Table 2.
\begin{table}[htb]
\centering
\begin{tabular}{cccccc}
\hline
$\sqrt{s}$ & $P_L$ & $\bar P_L$ &$\sigma_L$ (fb)&
$\frac{\delta\sigma}{\sigma}/\kappa$ & $\kappa_{\rm lim}$ (\%) \\ 
\hline
 250 & 0 & 0 & $56.4$ & 1.148 &  77.9  \\
     & $-0.8$ & 0 & $71.9$ & 1.094 & 72.4  \\
     & $-0.8$ & $+0.3$ &$91.3 $ & 1.090 & 64.5  \\
 350 & 0 & 0 & $56.6$ & 0.313 &  286  \\
     & $-0.8$ & 0 & $86.0$ & 0.318 & 228   \\
     & $-0.8$ & $+0.3$ & $111$ & 0.318 & 201   \\
 500 & 0 & 0 & $86.6 $ &$ 0.254$ & $352 $   \\
     & $-0.8$ & 0 &  $149 $   &$ 0.275$ &$ 248$  \\
     & $-0.8$ & $+0.3$ &  $193 $  &$ 0.277$ &$ 216$ \\
 1000 & 0 & 0 &  $ 214 $  & $ 0.296$ & $ 302$   \\
     & $-0.8$ & 0 & $384 $  &$  0.299$ &$ 224 $ \\
     & $-0.8$ & $+0.3$ &  $499 $  &$  0.299$&$ 196$ \\
\hline
\end{tabular}
\caption{The cross section $\sigma_L$ for $H\nu\bar\nu$ production, the fractional change in the cross 
section $\sigma_L$ for unit
value of $\kappa$ and the sensitivity to $\kappa$ for various c.m.
energies $\sqrt{s}$ and combinations of $e^-$ and $e^+$ longitudinal
polarizations $P_L$ and $\bar P_L$, respectively. An integrated
luminosity of 2 ab$^{-1}$ is assumed.}
\end{table}
Since for values of $\sqrt{s}$ less than or reaching 500 GeV the $HZ$
production process dominates, the results are similar to those shown in
Table 1. For higher energies, however, even though the cross section for
$WW$ fusion dominates, the sensitivity of this process to $\kappa$ is
somewhat less, since there is a partial cancellation between the two
mechanisms, the dependence on $\kappa$ of the $HZ$ production cross
section being negative.
Thus, while longitudinal polarization does help, the $HZ$ production
channel with $Z$ decay into visible channels is still better for the
measurement of $\kappa$, as compared to the $\nu \bar \nu$ channel.
However, it is possible to combine data from these two processes at the
ILC to get a sensitivity which is better than individual sensitivities
of the two processes.

\section{Conclusions and Discussion} 
The determination of the Higgs self-coupling is an
important issue for the confirmation of the SM or establishing the
veracity of a possible extension of the SM. Since the cross section for
Higgs pair production, which could measure the triple Higgs coupling
directly, is small, an indirect measurement through the process of
associated single Higgs production may have an upper hand. Following
an earlier suggestion \cite{McCullough:2013rea} 
for such a measurement in the process $e^+e^- \to
ZH$, where $\lambda_3$ could be measured through its one-loop
contribution, we have investigated the effect of beam polarization on
such a measurement. Again, it is assumed that the only source of
additional contribution is the $\kappa$ term, and we work in the lowest order
in $\kappa$. Our calculation shows that the fractional change in the
cross section arising from one-loop contribution of the triple-Higgs
coupling itself does not depend on polarization. Nevertheless, since
the cross section does improve with polarization, we find that 
reasonable longitudinal beam polarizations of
$-0.8$ and $+0.3$ as foreseen for the ILC leads to an
improvement of about 19\% in the sensitivity to $\kappa$. Even if only
the electron beam is polarized, there is still an improvement in the
sensitivity by about 8\%. 

This also implies that if polarized beams are used, the same
sensitivity as obtained with unpolarized beams can be achieved with a
33\% lower luminosity of 1.34 ab$^{-1}$.
We have also calculated the effect of beam polarization for a final
state $H\nu\bar \nu$ which gets contribution from $ZH$ production with
$Z$ decaying into $\nu\bar \nu$ as well as from $WW$ fusion.
Here we do find that the fractional change in the cross section with
$\kappa$ depends on the degree of beam polarization. However, for
somewhat higher energies, where the total cross section is higher, 
this change has opposite signs for the $ZH$ and $WW$ fusion contributions,
and tends to reduce the sensitivity.
We find an improvement of 16\% in the sensitivity as compared to the
case of unpolarized beams for a c.m. energy of 250 GeV. 
The data from $H\nu\bar \nu$ and $H\mu\bar \mu$ final states may be
combined to get a much better sensitivity. Again, the sign-reversed
polarization combination $(+0.8,-0.3$) gives a worse sensitivity.

We now discuss our results in the context of other recent relevant 
studies. 
As mentioned earlier, it is expected that HL-LHC, at the end of its run will be
able to determine $\kappa$ with an accuracy of about 50\%. There have been
various studies suggesting how these results could be combined with results
from the planned $e^+e^-$ experiments, see, for example,
\cite{deBlas:2019rxi}. 
However, to be able to compare our
results, we restrict ourselves to results which could be obtained separately
from $e^+e^-$ experiments. 

The  idea
of using single-Higgs production $e^+e^- \to HZ$ 
was suggested by \cite{McCullough:2013rea}. 
This work did not consider the effect of beam polarization, which we 
have included.
We have also included in the study the process $e^+e^- \to H\nu\bar \nu$, not
included in \cite{McCullough:2013rea}, and which gets
additional contribution from $WW$ fusion. 

%We have given analytic expressions, which can be used in different contexts,
%and also listed cross sections in the presence of electron and/or positron
%beam polarization for various energies in Tables 1 and 2.
%We have also pointed out that though 
%the relative contribution of the one-loop diagram with the 
%triple-Higgs coupling is not improved by polarization, the actual total cross
%section is higher for our choice of polarization, which improves the
%sensitivity.

In the context of ILC, in a recent report
it was estimated that Higgs pair production at $\sqrt{s} = 500$ GeV and an
integrated luminosity of 4 ab$^{-1}$, $\kappa$ could be determined to an
accuracy  of 27\% using $e^+e^- \to ZH$ \cite{Bambade:2019fyw}
with a run involving certain mix of polarization.
Using
$e^+e^- \to ZH$, the accuracy expected is 40\% for the full run of ILC in the
energy range 250 GeV to 500 GeV 
\cite{Bambade:2019fyw}, with most of the contribution from the Higgs 
loop stated to be coming below 350 GeV. The report does not mention the
process $e^+e^- \to H\nu\bar \nu$, however. The results in
\cite{Bambade:2019fyw} are obtained using detailed simulation, which we have
not attempted. Again, they report on results using a certain scheduling which
includes runs with all beam polarization combinations, with view to
optimizing sensitivity to several EFT couplings. 
If we do a simple-minded
extrapolation of our results from Tables 1 and 2 to an integrated luminosity
of 4 ab$^{-1}$ for a run at $\sqrt{s} = 250$ GeV
a combination of the $ZH$ and $H\nu\bar\nu$ data would lead us to
an accuracy of about 37\% for unpolarized beams, and about 30\% for both
beams polarized in the combination $(-0.8,+0.3)$. It thus seems that from
the point of view of triple-Higgs couplings, the accuracy mentioned in 
\cite{Bambade:2019fyw} can be achieved, or perhaps even superceded with the
use of the same integrated luminosity without an energy upgrade.
%with only electron beam polarized with the same assumptions is 34\%, a fair
%improvement over the unpolarized case.  

At the CEPC, with a precision of 0.5\% on the measurement of the $ZH$ cross
section at 240 GeV and an integrated luminosity of 5.6 ab$^{-1}$, 
$\kappa$ can be constrained to 35\% \cite{An:2018dwb}. This is an update on
the original proposal \cite{TLEP} alluded to by \cite{McCullough:2013rea}, which we
mentioned earlier. As CEPC is not
designed to have polarized beams, our suggestion for the use of polarization
does not have a direct
impact on this result. However, a similar capability would be possible at
ILC with polarized beams with a lower luminosity of about 5 ab$^{-1}$.
Similarly, at FCC-ee, the capability of measurement of $\kappa$ is expected to
be similar to that at CEPC \cite{Baglio:2019gmz}.

Proposals at CLIC, on the other hand, have considered the possibility of only
the electron beam having 80\% polarization.  
In a recent study in Ref.\cite{Roloff:2019crr}  
it is shown that with an integrated luminosity of 5 ab$^{-1}$ and 
$\sqrt{s}=3$ TeV, CLIC will be able to measure trilinear Higgs coupling 
with an uncertainty of $-7$\% 
and 11\% with C.L. 68\%. Single Higgs production at this high energy would
not be significant. On the other hand, the integrated luminosity proposed
for CLIC to operate at lower energies of 350 or 380 GeV is low. However, our
study allows us to anticipate that if CLIC is operated at lower energy with
longitudinal polarization for a sufficient length of time, $ZH$ production
would allow a significant limit to be put on $\kappa$.

Unlike our present work which discusses only the total cross sections, 
better sensitivity might perhaps be obtained with use of differential
decay distributions of charged leptons, and needs to be investigated in
detail. 
A significant work points out the possibility of measuring $\kappa$ directly
from single Higgs production making use of a certain T-odd kinematic
asymmetry of decay leptons, 
which is not affected by anomalous tree-level couplings
\cite{Nakamura:2018bli}. 
This work estimates a limit on $\kappa$ of order 1 for an
integrated luminosity of 30 fb$^{-1}$ making use of beam polarization as
well as $\tau$ polarization in $Z$ decay.

The discussion here has included only statistical uncertainties. 
It is appropriate to mention possible systematic uncertainties
which may be anticipated. These would be mainly related to $b$ tagging
when using the decay $H \to b \bar b$ for Higgs detection, luminosity
measurement, and polarization measurement. The systematic errors in
luminosity and polarization measurements are each estimated to be 0.1\%
\cite{Asner,Bambade:2019fyw}.
The systematic errors due to $b$-tagging
efficiency are estimated to be 0.3\%$\sqrt{0.250/L}$, where $L$ is the 
integrated luminosity in ab$^{-1}$ \cite{Bambade:2019fyw}. 
Thus the systematic uncertainty
of the ILC option we consider would be within about 0.1\%.
It is thus clear that the systematic uncertainties are much
lower than the statistical uncertainties in the measurement of 
the Higgs coupling, and may be ignored.

It is hoped that we have made a case for exploring
seriously the possibility of implementing longitudinal polarization at
future $e^+e^-$ colliders. While polarization of both electron and positron
beams would be extremely useful, even a high electron beam polarization
in the absence of positron polarization would serve a useful purpose.

We have assumed for simplicity that a run with polarized beams would
collect the full integrated luminosity of 2 ab$^{-1}$. 
Also, we have made a simplified assumption regarding the detection
efficiencies. A more realistic simulation, including proper isolation
cuts and detector efficiencies, would be needed to determine the actual
improvement in the sensitivity in a practical situation. However, our
results do provide a reasonable first estimate of the advantage of beam
polarization.

As mentioned earlier, there may be a BSM tree-level contribution to the
process through an anomalous $ZZH$ coupling, and this will have to be
subtracted before extracting the one-loop effect of $\kappa$ discussed
here. One way, is to use results from two different
energies, as was discussed in ref. \cite{McCullough:2013rea}. Another
possibility of separating the two contributions would be the use
of different kinematic 
correlations in the final state arising from $Z$ decay.
As mentioned earlier, the work of ref. \cite{Nakamura:2018bli} makes uses of 
such a kinematic asymmetry which does not get contribution from tree-level
$ZZH$ couplings.
Our treatment assumes that the production process proceeds through a
virtual $Z$. There would also be a contribution from a virtual $\gamma$
state at tree level through an anomalous $\gamma ZH$ coupling, or 
at one-loop level, though not from the triple-Higgs coupling. 
Though we have not taken into account this
possibility, it is possible, as shown in \cite{Rindani:2009pb}, to use
either more than one beam polarization combination or more than one
polar-angle cut-off to determine separately the $Z^*$ contribution.
Ref. \cite{Rindani:2009pb} also shows how transverse beam polarization,
if available, may be used to achieve the same purpose.

\noindent{\bf Acknowledgement} SDR acknowledges support from the Department of Science and Technology,
India, under the J.C. Bose National Fellowship programme, Grant No.
SR/SB/JCB-42/2009.

\end{document}